\documentclass[12pt]{article}
\usepackage{times}
\usepackage{geometry}
\usepackage{graphicx}

\usepackage[utf8]{inputenc}
\usepackage{enumitem,amssymb}
\usepackage{ragged2e}
\usepackage{afterpage}
\newlist{thematic}{itemize}{8}
\setlist[thematic]{label=$\square$}
\usepackage{pifont}
\usepackage{natbib}
\def\la{\mathrel{\mathchoice {\vcenter{\offinterlineskip\halign{\hfil
$\displaystyle##$\hfil\cr<\cr\sim\cr}}}
{\vcenter{\offinterlineskip\halign{\hfil$\textstyle##$\hfil\cr
<\cr\sim\cr}}}
{\vcenter{\offinterlineskip\halign{\hfil$\scriptstyle##$\hfil\cr
<\cr\sim\cr}}}
{\vcenter{\offinterlineskip\halign{\hfil$\scriptscriptstyle##$\hfil\cr
<\cr\sim\cr}}}}}

\newcommand*\aap{A\&A}

\newcommand*\apj{ApJ}
\newcommand*\apjl{ApJ}

\newcommand*\mnras{MNRAS}

\newcommand*\nat{Nature}

\begin{document}
\raggedright
\huge
Astro2020 Science White Paper \linebreak

Extremely metal-poor stars:\\ the need for UV spectra \linebreak
\normalsize

\noindent \textbf{Thematic Areas:} \hspace*{60pt}  $\square$ Star and Planet Formation \hspace*{20pt}\linebreak
 $\square$ Cosmology and Fundamental Physics \linebreak
  $\square$  Stars and Stellar Evolution \hspace*{1pt} $\square$ Resolved Stellar Populations and their Environments \hspace*{40pt} \linebreak
   
\textbf{Principal Author:}

Name:	Piercarlo Bonifacio
 \linebreak						
Institution: GEPI, Observatoire de Paris, Universit{\'e} PSL, CNRS 
 \linebreak
Email: Piercarlo.Bonifacio@observatoiredeparis.psl.eu 
 \linebreak
Phone:  
 \linebreak

\textbf{Co-authors:} Elisabetta Caffau
 \linebreak						
Institution: GEPI, Observatoire de Paris, Universit{\'e} PSL, CNRS 

\textbf{Co-authors:} Monique Spite
 \linebreak						
Institution: GEPI, Observatoire de Paris, Universit{\'e} PSL, CNRS 

%\textbf{Abstract  (optional):}

%Do we need an abstract?

\pagebreak
\section{Introduction}
Our current understanding of the chemical evolution of the Universe is
that a first generation of stars was formed out of primordial material,
completely devoid of metals (Pop III stars). This first population of
stars comprised massive stars that exploded as supernovae disseminating
the metals they synthesised in the interstellar medium. These
massive stars are long dead and cannot be observed in the local Universe.
Among very metal poor stars (metallicity below -2.0) we expect
to find the direct descendants of these pristine metal factories.
The chemical composition of these stars provides us indirect information
on the nature of the Pop III stars, their masses, luminosities and mode
of explosion. The constraints are stronger if the chemical inventory
is more complete, more chemical elements and isotopic ratios are
measured for each star. Unfortunately the lower the metallicity of the
star, the weaker the lines. Access to the space UV spectral range gives
us crucial supplementary information. To start with, it allows access to
some very strong Fe lines that may allow to measure the abundance of this
element in stars for which this was not possible from the ground-accessible
UV spectra. The number of such stars is steadily increasing. Next the UV range
allows us to measure elements that cannot be measured from ground-based spectra like
P, Ge, As, Se, Cd, Te, Lu, Os, Ir, Pt, Au. In addition it is fundamental
for measuring other elements that can be accessed from earth, but with great
difficulty, like C, S, Cu, Zn, Pb. The Hubble space telescope, with its limited
collecting power made this possible only for very few stars. Old metal poor
stars are cool, of spectral types F,G,K, and their UV flux is low.
The availability of a UV high resolution spectrograph fed by a large area space 
telescope  will open an unprecedented window on the early evolution of
our Galaxy.

\section{Carbonicity: a path to understand star formation at very low metallicity ?}

Star formation in  the pristine Universe takes place under conditions very
different from what happens in the present day Universe.
As a gas-cloud collapses its temperature increases, and  its pressure tends
to counter-balance the gravitational pull. In present day gas there are two 
main mechanisms that allow a  contracting gas-cloud to cool, while collapsing, thus
ensuring that the collapse may continue: collisional excitation and radiative
recombination of atomic and molecular levels and collisional heating of dust particles
with successive emission of far-IR photons. In both cases the produced photons
can escape the collapsing cloud without further interacting with the gas,
thus effectively subtracting energy from the system and allowing the collapse to continue.
In a gas that is totally devoid of metals, there is no dust and the available
atoms and molecules (H, $^2$H, $^3$He, $^4$He, $^7$Li, H$_2$, LiH) either
do not have low-lying levels that can easily be excited collisionally or are
too rare to provide an efficient source of cooling.
From these simple physical considerations one expects that the first
generation of stars would be made only by massive stars, that result from
the collapse of clouds that are massive enough that the pull of gravity
cannot be counter-balanced by the increase of temperature during the collapse. 
Such stars would be long dead and not observable in the local Universe. 
Only low-mass stars have long enough lifetimes, that, if formed in the first generations
of stars, would still be shining today.
The discovery of the first star with [Fe/H] $<-5$, HE\,0107-5240 \citep{Christlieb02}, came
as a big surprise, because the record deficiency in Fe was accompanied by a large
overabundance in C, N (later it was clear that also O is largely overabundant
in this star \citealt{Bessel04}).
The review by \citet{Palla03} describes very well the state-of-the art of our understanding
of star formation at very low metallicity. 
Most of the community assumed that the peculiar chemical composition
of HE\,0107-5240 was the composition of the gas cloud out of which it was formed
 and was the result of
enrichment of one or several SNe \citep{Umeda03,Bonifacio03,Limongi03}.
The formation of low-mass stars from such gas   
was explored \citep{Schneider03}.

In a very influential paper  \citet{BL03} showed that the low lying levels
of C{\sc II} and O{\sc I} can provide a very efficient cooling mechanism,
through collisional excitation and radiative recombination, provided the
carbon or oxygen abundance is at least 0.01\% to 0.1\% of that found in the Sun.
A condition satisfied by HE\,0107-5240.
The peculiar chemical composition would then not be surprising, but the result of the fact
that such a chemical composition is necessary to allow the formation of low-mass stars.
It was thus not a surprise that when in the following years other stars of this
very low metallicity were found they all turned out to be highly enhanced
in carbon  \citep{Frebel05,Norris07}.
For several years it was generally accepted that below [Fe/H]$\la -4.5$
one should find only C-enhanced stars. In spite of the fact that a part
of the community privileged an interpretation of the C-enhancement in these stars as the 
result of mass-transfer from a former AGB companion \citep{Suda,Lau,Cruz},
a line of thought that has been reinforced by the recent discovery 
by \citet{Arentsen} that HE\,0107-5240 is, after all, a binary system.
Nevertheless the real shock in the field came by the discovery
of SDSS\,J1029+1729 by \citet{Caffau11} a star with [Fe/H]$\approx -5$ but
no evidence of carbon enhancement. Showing that low-mass
stars can be found even below the minimum amount of C and O required by
the \citet{BL03} scenario. This finding has  recently
been reinforced by the discovery of   Pristine\_221.8781+9.7844
by \citet{Starkenburg}, a star with similar properties to those
of  SDSS\,J1029+1729, making it unlikely that these stars are the
result of some highly contrived evolutionary scenario.

Nevertheless the supporters of the need for for C enhancement for the
formation of low-mass stars argue that 
SDSS\,J1029+1729 and  Pristine\_221.8781+9.7844
are in reality C-enhanced, simply the C abundance cannot be measured
because of the weakness of the G-band in these warm Turn-Off (TO) and Sub Giant (SG) stars
\citep[e.g][]{Placco}.
This claim is very difficult to confute, in spite of the fact that
the claim of C-enhancement in the TO stars G64-12 and G64-37 by \citet{Placco},
based on a 1D analysis of the G-band, is not 
supported by the 3D-NLTE analysis of IR atomic lines in these
stars by \citet{Amarsi}.
The reason is that the TO stars, based on evolutionary time-scales, are certainly much more
numerous than cooler giant stars and, at the same time, considerably brighter
than cooler MS stars. It is thus likely that other examples of carbon-normal
extremely metal-poor stars will be found among TO stars, where the G-band is 
weaker, for any given C abundance, with respect to what observed in cooler stars.
The IR C atomic lines observed by \citet{Amarsi} weaken with decreasing C abundance
to a level where they become not observable in stars like
SDSS\,J1029+1729 and  Pristine\_221.8781+9.7844 with current and future observational
facilities.
If we wish to measure a C abundance in SDSS\,J1029+1729 and  Pristine\_221.8781+9.7844 and
their siblings, rather than mere upper limits,  we have to turn to the vacuum UV.  
The C{\sc I} line at 193.09\,nm is so strong that it would be measurable
in SDSS\,J1029+1729 and  Pristine\_221.8781+9.7844, even if they had [C/Fe]=0.0
\citep[e.g.][]{Spite17}. Currently the instrumentation on board of HST cannot
tackle these faint cool stars (V$\approx 17$). 
A space telescope with a large collecting
area (e.g. LUVOIR) equipped with a high-resolution UV spectrograph
should give us a definite answer on the question:
do carbon-normal stars exist with [Fe/H] $< -4.5$ ?

%-------------------------------------- 
   \begin{figure}
   \centering
\resizebox{!}{7cm}{
	   \includegraphics[clip=true]{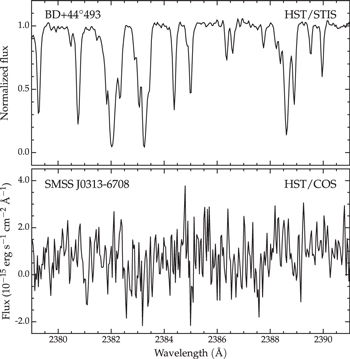}
           \includegraphics[clip=true]{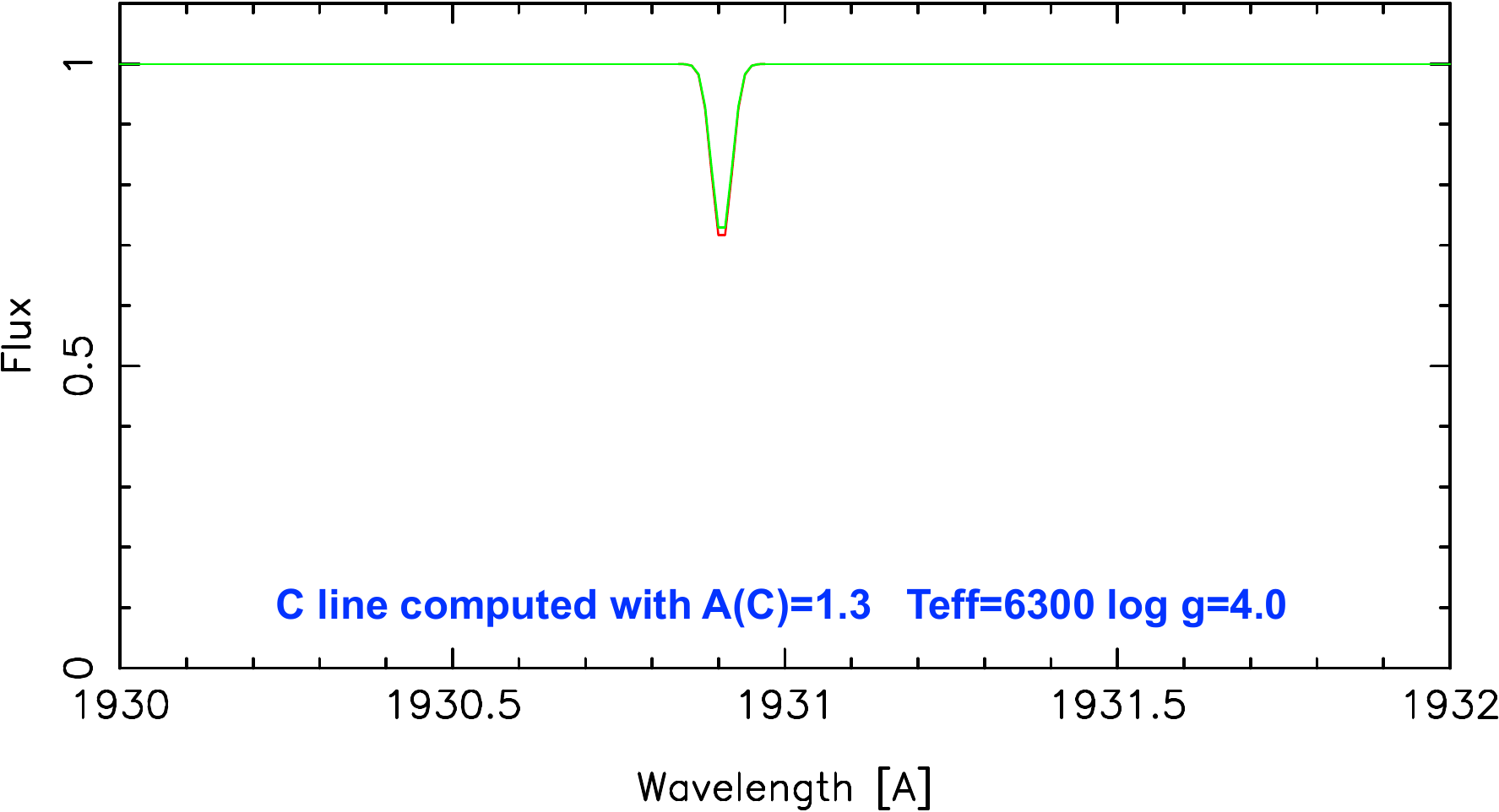}}
   \caption{Left: the HST COS spectrum of SMSS\,0313-6708, compared to the HST STIS
spectrum of the metal-poor star BD +44  493 ([Fe/H]=--3.7) around the
strong Fe {\sc II} line at 2382 \AA\, \citep{RoedererKS}. Right: synthetic computation of the UV C\,I line
for a turn-off stars with A(C)=1.3, correspondent to about [C/Fe]=--7.}
              \label{fig:Fe}%
    \end{figure}
%-------------------------------------- 

\section{What is the iron abundance ?}

As we find stars that are more and more metal-poor it occurs
that we are not able to measure any Fe line from optical
or near-UV spectra. Currently there are four stars for which
no iron line could be measured:
SMSS\,0313-6708 ([Fe/H] $<-7.10$ \citealt{Keller}),
SDSS\,J1035+0641 ([Fe/H] $<-5.20$ \citealt{Bonifacio18}),
SDSS\,J0023+0307 ([Fe/H] $<-6.0 $ \citealt{Aguado18b}),
SDSS\,J0815+4729   ([Fe/H] $<-5.8 $ \citealt{Aguado18a}). 
In these stars only Ca and C can be measured.
We expect several more to be discovered by the on-going and planned surveys
and it would be of paramount importance to determine their
Fe abundance.
\citet{RoedererKS} integrated with HST COS 32\,114 s, spread over 10 orbits,
on  SMSS\,0313-6708, 
but was unable to detect any measurable line. This example illustrates
very well the need for a larger collecting power than HST (e.g. LUVOIR) 
and spectral
performances, at least at the level of COS.

\section{Phosphorus a shy element}

\begin{figure}
   \centering
\resizebox{!}{4.30cm}{
	   \includegraphics[clip=true]{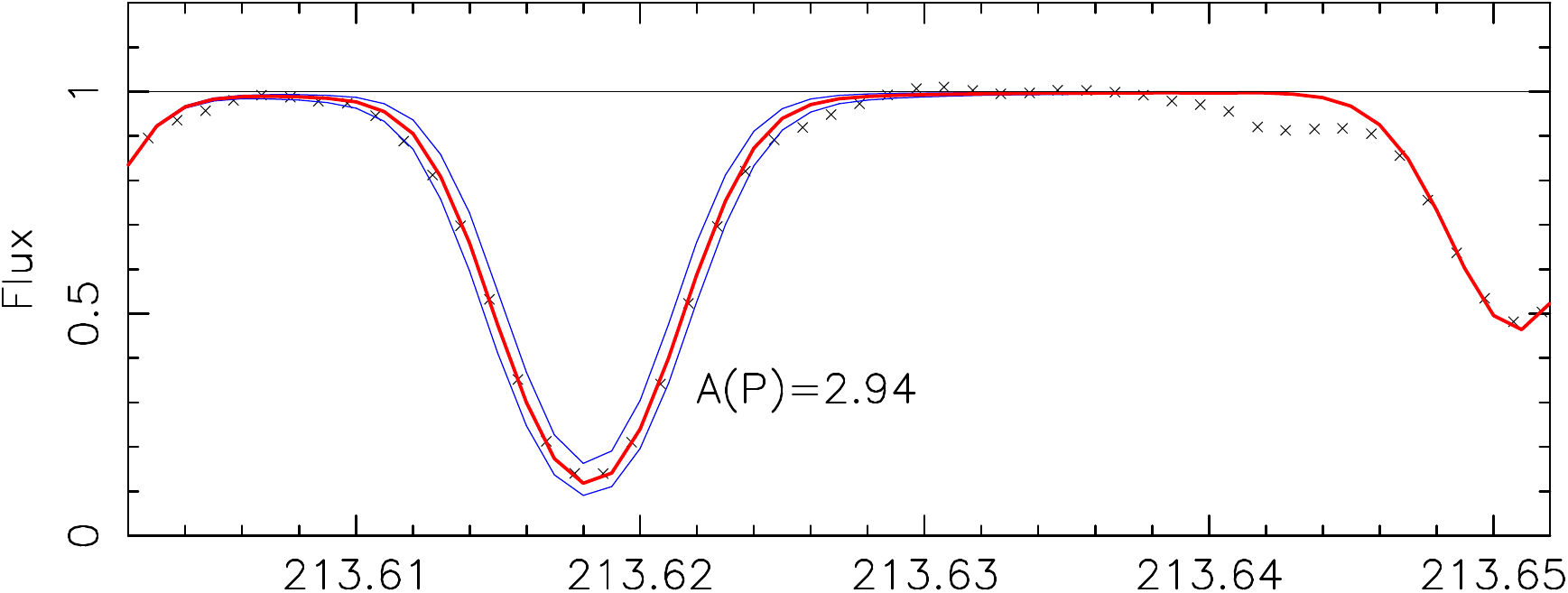}
}

\resizebox{!}{5.0cm}{
	   \includegraphics[clip=true]{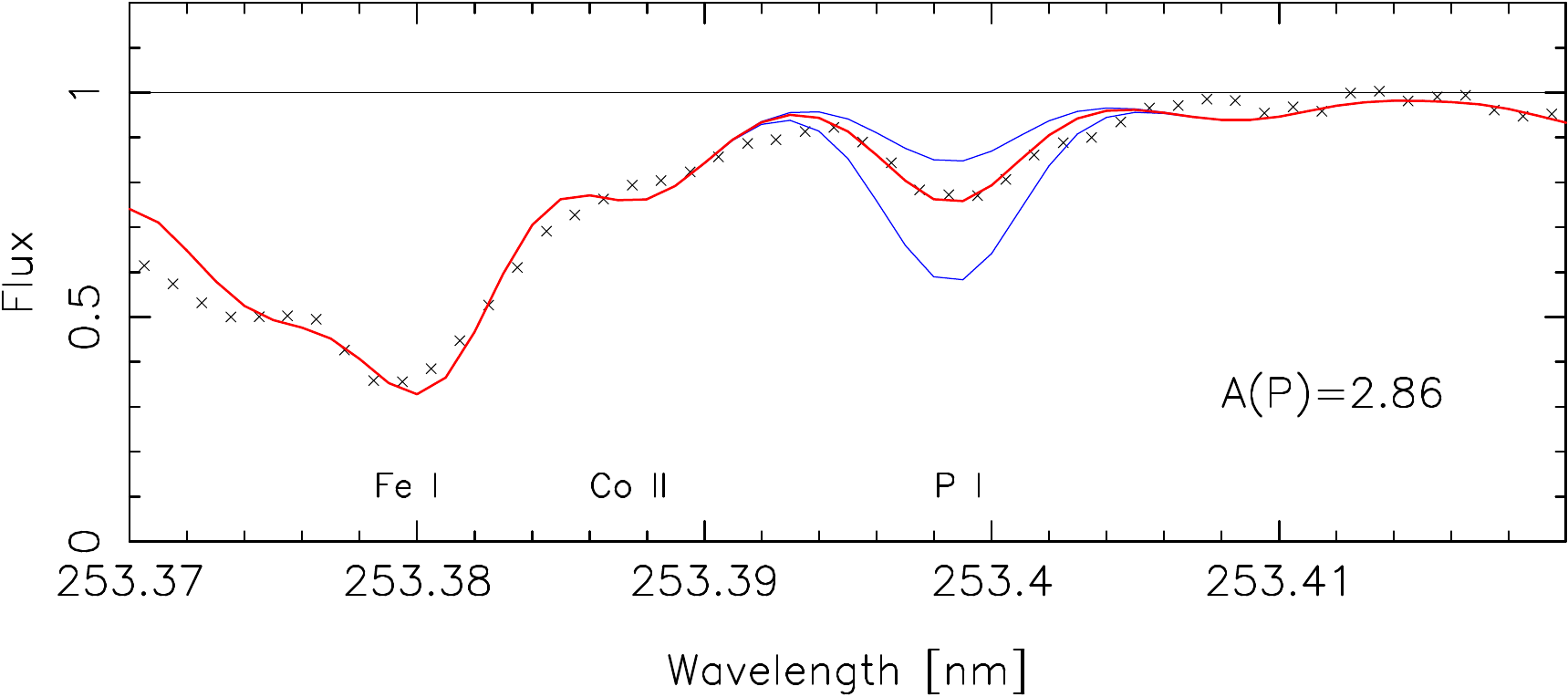}
}
   \caption{From figure\,6 of \citet{Spite17}: two P\,I lines in the HST spectrum of HD\,84937 (black crosses), compared to the synthetic spectra
            (solid red and blue lines) in the range of four P\,I lines.}
              \label{fig:P}%
    \end{figure}

Phosphorus is a relatively abundant element in the Universe.
Its two closest elements in the periodic table of elements (Si and S) are well investigated
while P remains mysterious because only few P lines (in the infra-red and in the UV) can be detected in a stellar spectrum.
The P\,I lines in the infra-red are relatively weak and can provide the P abundance in stars with
a P abundance from over-solar down to about 1/10 the solar P \citep[see e.g.][]{pCaffau11}. 
Some P\,I lines in the UV allow to derive P abundances at metal-poor regime
\citep[see][]{Roederer14} so to allow the investigation of the Galactic evolution of phosphorus.
\citet{Roederer14} investigated ten P\,I lines in the UV at metal-poor regime, but only few were useful to derive abundances.
\citet{Spite17} selected five P\,I UV lines (two are shown in Fig.\,\ref{fig:P}) to derive the P abundance of the metal-poor star HD\,84937 and obtained a
good agreement in A(P) from the lines she investigated.
A high-resolution UV spectrograph with
large effective area,
able to investigate metal-poor stars will allow to finally understand the Galactic chemical evolution of phosphorus.

%-------------------------------------- 
%-------------------------------------- 

\section{Heavy elements}

\begin{figure}
   \centering
\resizebox{!}{4.30cm}{
	   \includegraphics[clip=true]{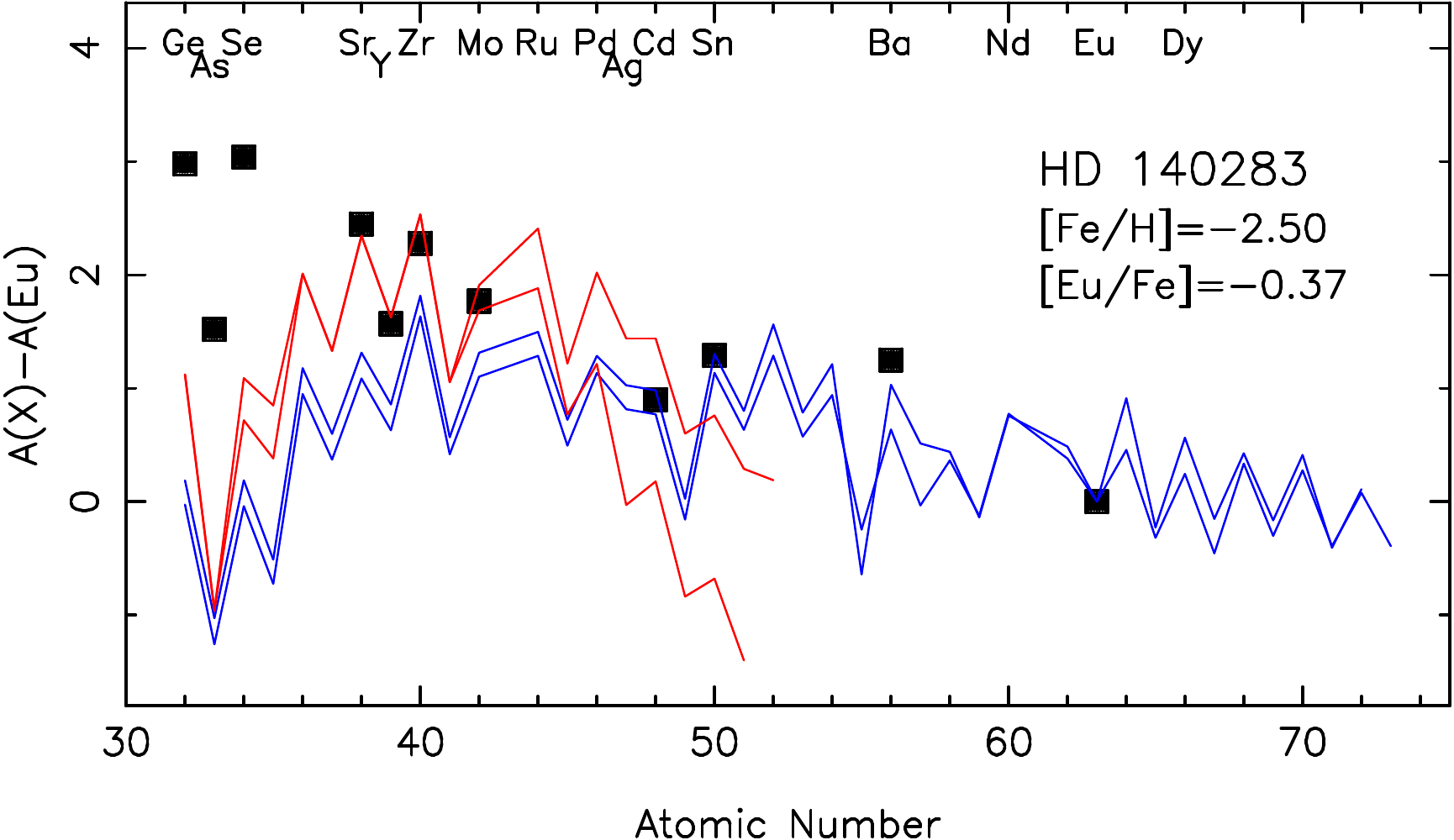}
}
   \caption{The chemical pattern of HD\,140283 compared to theoretical models.}
              \label{fig:heavy}%
    \end{figure}

It is believed that in the very old, very metal-poor stars, 
born in the first Gyr of our Galaxy, most of the neutron capture elements were produced by the main 
r-process in neutron star mergers \citep[e.g.][]{Pian}. 
But this process is clearly not able to explain the abundance of 
all the heavy elements and in particular the abundance of the lightest of these elements like Ge, As, Se
\citep[e.g.][]{Spite18}.  
It is very important to compare the complete distribution of the abundance of the heavy elements 
to the predictions of the different processes: main-r- process, weak-r processes, 
i(intermediate) process, neutron-rich neutrino winds in core-collapse supernovae etc... 
Up to now no process is able to explain the pattern of the heavy elements in metal-poor stars (see Fig.\,\ref{fig:heavy}). 
Many neutron capture elements are impossible to detect in the spectral range accessible 
to ground-based telescopes. 
Their abundance can be deduced only from the UV
lines between 190 and 300 nm (e.g. Ge, As, Se, Cd, and also Lu, Pt and
Au), hence a high-resolution UV spectrograph is required.

%-------------------------------------- 
%-------------------------------------- 

%    \vfill\eject
%\pagebreak
    \afterpage{\clearpage}

\bibliographystyle{aa}
\bibpunct{(}{)}{;}{a}{}{,} 

%\pagebreak
%\textbf{References}
{}

\end{document}